\documentclass[10pt,journal,compsoc]{IEEEtran}
\usepackage[nocompress]{cite}
\usepackage{amsmath,amssymb,amsfonts}
\usepackage{algorithmic}
\usepackage{algorithm}
\usepackage{graphicx}
\usepackage{textcomp}
\def\BibTeX{{\rm B\kern-.05em{\sc i\kern-.025em b}\kern-.08em
    T\kern-.1667em\lower.7ex\hbox{E}\kern-.125emX}}
\begin{document}

\title{Learning Slab Classes to Alleviate Memory Holes in Memcached}
\author{Devang~Jhabakh~Jai~and~Sudeep~Das}

\IEEEtitleabstractindextext{%
\begin{abstract}
We consider the problem of memory holes in slab allocators, where an item entered into memory occupies more memory than it actually requires due to a difference between the nearest larger slab class size and the size of the entered item. We solve this problem by using a greedy algorithm that analyses the pattern of the sizes of items previously entered into the memory and accordingly re-configuring the default slab classes to better suit the learned traffic pattern to minimize memory holes. Using this approach for a consistent data pattern, in our findings, has yielded significant reductions in memory wastage. We consider Memcached as it is one of the most widely used implementations of slab allocators today, and has native support to reconfigure its default slab classes.
\end{abstract}

}

\maketitle

\section{Introduction}
\IEEEPARstart{M}{emcached} uses a slab allocator to store and retrieve values stored in a cache. Memcached's slab allocator segregates items into different slab classes. Slab classes are composed of pages, each of which contains chunks of a fixed size for that given slab. The use of a slab allocator helps alleviate external fragmentation caused due to allocations and deallocations. However, due to the chunk sizes reserved for items being of a fixed set of sizes, the memory allocated for an item is often more than what is used by it - this leads to internal fragmentation, or memory holes.

When a key-value pair is inserted in Memcached, the actual memory required by this item is equal to the size of the key + value + miscellaneous data from Memcached’s internal data structures \cite{b1}. Due to internal fragmentation, the memory allocated for such an item is often more than the memory used by it. This is because the item sizes may not line up with the sizes of the Slab Classes as set by Memcached – by default starting at 96 bytes and geometrically scaling up by a factor of 1.25\textsuperscript{n} till the 1 MB page size limit, the size of one full page.

In our testing, we saw an average 10$\%$ wastage in memory due to internal fragmentation for log-normal traffic patterns. As Memcached instances scale up in size, this may lead to increasingly large magnitudes of waste. To solve this problem can help recover terabytes of usable memory for large scale implementations of Memcached.

\section{Definitions}

\subsection{Chunks}

A chunk is the smallest unit of storage in Memcached, and it holds exactly one item. The size of a chunk is decided on the basis of the slab class it belongs to.

\subsection{Pages}

Pages, of size 1MB, are used to store initialized chunks in memory. Memory is allocated one page at a time. Each page initializes all chunks of the same size.

\subsection{Slabs}

Slabs hold a collection of pages. Memcached organizes collections of pages into slabs that belong to different classes corresponding to object sizes. 

\subsection{Memory Hole}

Memory Hole is the term used to refer to the memory wasted as a result of internal fragmentation within a chunk.

\subsection{Problem Statement}

The problem of alleviating memory holes can be stated as follows: Given a distribution of the frequency of occurrence of an item for given item sizes, we aim to find the optimal set of slab chunk sizes such that memory holes due to internal fragmentation can be minimized, while keeping the number of slab classes constant.

\section{Related Work}

The analysis of factors like Memcached's latency\cite{b2}, hit rates\cite{b3}, power\cite{b4} and scalability\cite{b2},\cite{b5} have been the subject of past studies. On the other hand, work related to the alleviation of its memory holes is scarce. The developers of Memcached have, however, made adjustments to try and alleviate memory holes to some degree by allowing users to change the value of the default slab size growth factor of 1.25 using Memcached's startup options. However, lowering this growth factor to increase memory efficiency may come at the cost of significantly increasing the eviction rates for some classes.

To date, there exists no other work in the management of slab classes to solve the issue of internal fragmentation in Memcached. 

\section{Solution}

We used Memcached as an implementation of slab allocator, and baselined its internal fragmentation under various traffic patterns, specifically such that the probability distribution of frequency versus sizes follows a pattern that has a large number of values centered around a certain median size, as is characterized by the use of Memcached at Facebook \cite{b2}. We then devised an algorithm to emit allocator slab sizes that minimize internal fragmentation. We applied the generated parameters to Memcached, and measured fragmentation under conditions identical to the original baseline test. We aim to provide an empirical study of the effects of trying to use our algorithm to alleviate memory holes.

Our greedy algorithm is an implementation of the hill climbing algorithm that attempts to \textit{maximise} the efficiency of the usage of memory. However, we restate our objective as trying to \textit{minimise} memory holes. We believe that this is an ideal algorithm as it is capable of producing a global minimum each time, given that if there is scope for improvement, it must always lie in a neighboring slab class configuration. The algorithm takes the probability distribution as an input and then tries different configurations of slab chunk sizes using the hill climbing strategy until it finds an optimal configuration of chunk sizes that wastes the minimum amount of memory. This new configuration is returned and can now be adopted.

\newlength\myindent
\setlength\myindent{2em}
\newcommand\bindent{
  \begingroup
  \setlength{\itemindent}{\myindent}
  \addtolength{\algorithmicindent}{\myindent}
}
\newcommand\eindent{\endgroup}
\begin{algorithm}[H]
\begin{algorithmic}
\STATE $slabs$ = List of sizes of all currently initialized slab classes
\STATE $oldwaste$ = Find the Current Memory wasted
\STATE $count$ = 0
    \STATE \textbf{do}
    \bindent
        \STATE Temporarily move a randomly selected $slabs$'s chunk size up or down 1 byte
        \STATE $newwaste$ =  Find the new Current Memory waste
        \IF{$newwaste \leq oldwaste$}
            \STATE $newwaste = oldwaste$
            \STATE $count$ = 0
        \ELSE
            \STATE Reset the Slab chunk sizes to the old configuration
            \STATE $count = count + 1$
        \ENDIF
    \eindent
    \STATE \textbf{while} $count \leq 1000$ 
    \STATE \textbf{return} $slabs$
\end{algorithmic}
\caption{Hill climbing algorithm}
\end{algorithm}

To use the returned configuration, one may use the "-o slab\_sizes" startup option with Memcached, and pass the desired list of slab chunk sizes as arguments. 

The aforementioned algorithm was run on a test server with traffic patterns that followed a log-normal probability distribution. Upon entering over 1 million items for each distribution, we observed the following results concerning reduction in memory wastage for some different sets of means and standard deviations of the log-normal distribution, where ''Old Configuration'' and ''New Configuration'' represent the state of the Memcached instances before and after the algorithm is run, respectively:
\vspace{20px}
\section{Algorithm testing and results}

\subsection{$\mu$ = 518 bytes, $\sigma$ = 10.5 bytes}

\begin{figure}[H]
\centerline{\includegraphics[scale = 1]{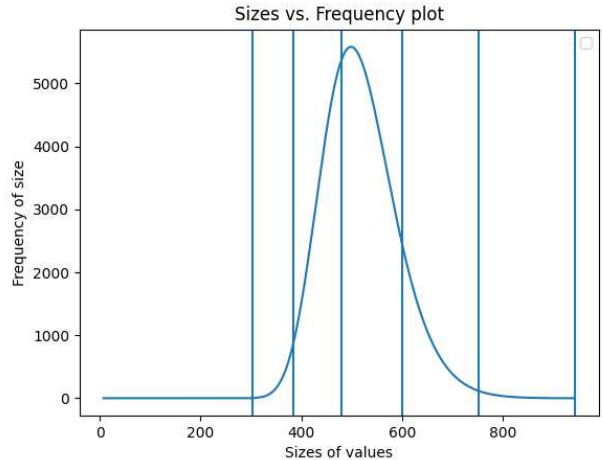}}
\caption{Old configuration of Slab Class chunk sizes. Each vertical line represents a slab class' chunk size}
\label{fig1.1}
\end{figure}

\begin{table}[H]
\caption{$\mu$ = 518 bytes, $\sigma$ = 10.5 bytes}
\begin{center}
\scalebox{0.86}{
\begin{tabular}{|c|c|c|c|}
\hline
\textbf{Measurement Metric}&{\textbf{Old Configuration}}&{\textbf{New Configuration}}\\
\hline
Available Chunk Sizes & [304,384,480,600,752,944] & [461,510,557,614,702,943]\\
\hline
Memory wasted (bytes) & 62,013,552 & 32,809,986\\
\hline
\end{tabular}}
\label{tab1}
\end{center}
\end{table}

\begin{figure}[H]
\centerline{\includegraphics[scale = 1]{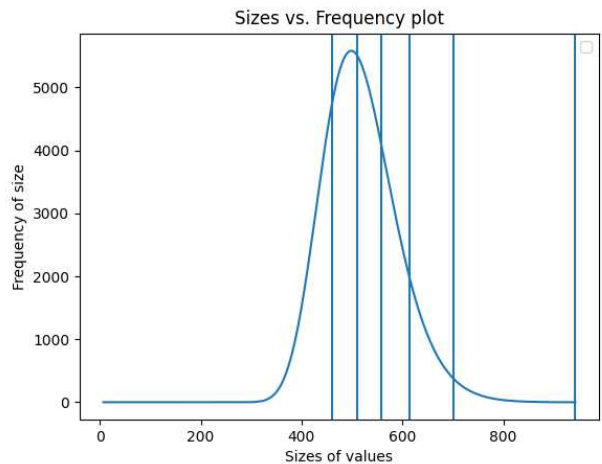}}
\caption{New configuration of Slab Class chunk sizes. Each vertical line represents a slab class' chunk size}
\label{fig1.2}
\end{figure}

{\textbf{The percentage of wasted memory recovered in the above case is 47.09$\%$}}
\\
\\
\\

\subsection{$\mu$ = 1210 bytes, $\sigma$ = 15.8 bytes}
\begin{figure}[H]
\centerline{\includegraphics[scale = 1.0]{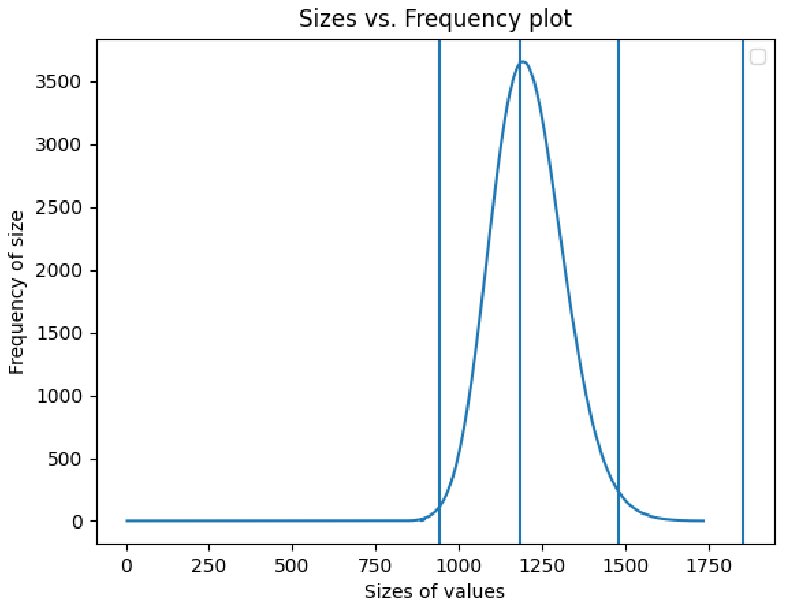}}
\caption{Old configuration of Slab Class chunk sizes. Each vertical line represents a slab class' chunk size}
\label{fig2.1}
\end{figure}

\begin{table}[H]
\caption{$\mu$ = 1210 bytes, $\sigma$ = 15.8 bytes}
\begin{center}
\scalebox{0.9}{
\begin{tabular}{|c|c|c|c|}
\hline
\textbf{Measurement Metric}&{\textbf{Old Configuration}}&{\textbf{New Configuration}}\\
\hline
Available Chunk Sizes & [944,1184,1480,1856] & [1173,1280,1414,1735]\\
\hline
Memory wasted (bytes) & 147,403,935  & 74,979,930\\
\hline
\end{tabular}}
\label{tab2}
\end{center}
\end{table}

\begin{figure}[H]
\centerline{\includegraphics[scale = 1.0]{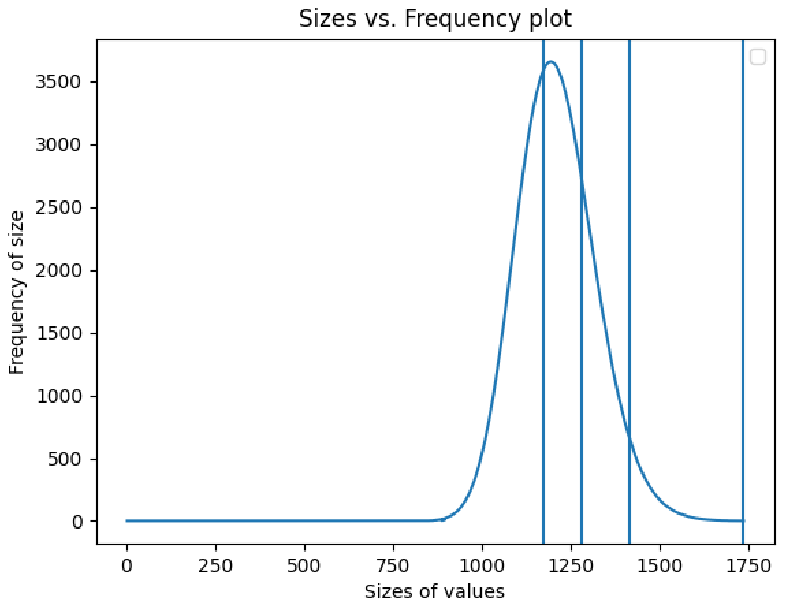}}
\caption{New configuration of Slab Class chunk sizes. Each vertical line represents a slab class' chunk size}
\label{fig2.2}
\end{figure}

{\textbf{The percentage of wasted memory recovered in the above case is 49.13$\%$}}
\\
\\
\\

\subsection{$\mu$ = 2109 bytes, $\sigma$ = 16.6 bytes}

\begin{figure}[H]
\centerline{\includegraphics[scale = 1.0]{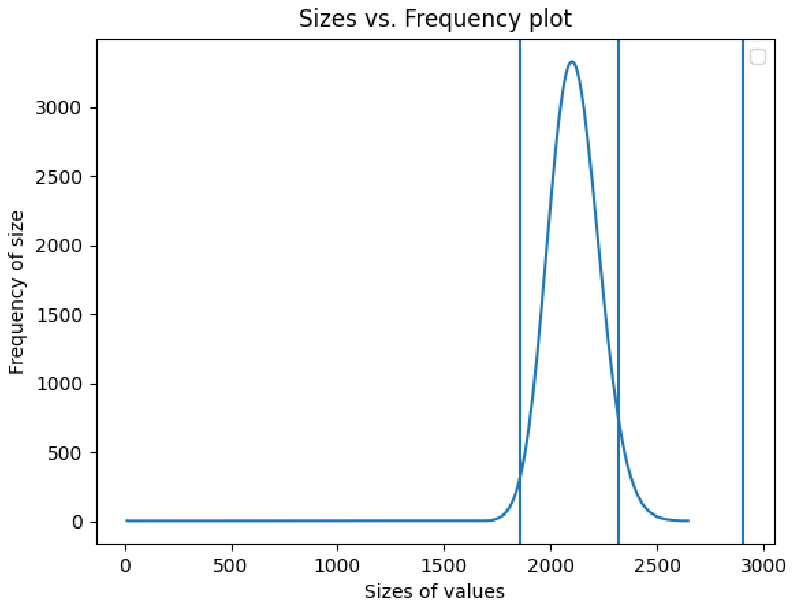}}
\caption{Old configuration of Slab Class chunk sizes. Each vertical line represents a slab class' chunk size}
\label{fig3.1}
\end{figure}

\begin{table}[H]
\caption{$\mu$ = 2109 bytes, $\sigma$ = 16.6 bytes}
\begin{center}
\scalebox{0.9}{
\begin{tabular}{|c|c|c|c|}
\hline
\textbf{Measurement Metric}&{\textbf{Old Configuration}}&{\textbf{New Configuration}}\\
\hline
Available Chunk Sizes & [1856,2320,2904] & [2120,2287,2643]\\
\hline
Memory wasted (bytes) & 230,144,462 & 111,980,981\\
\hline
\end{tabular}}
\label{tab3}
\end{center}
\end{table}

\begin{figure}[H]
\centerline{\includegraphics[scale = 1.0]{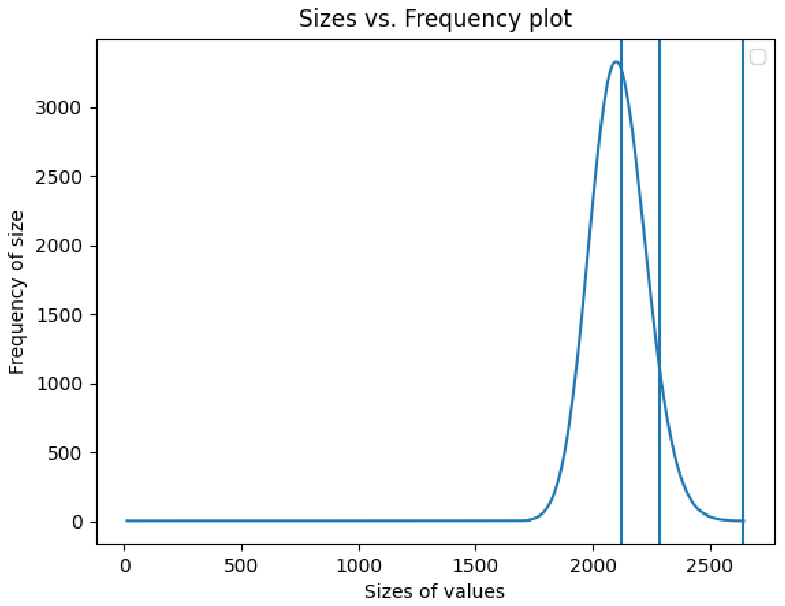}}
\caption{New configuration of Slab Class chunk sizes. Each vertical line represents a slab class' chunk size}
\label{fig3.2}
\end{figure}

{\textbf{The percentage of wasted memory recovered in the above case is 51.34$\%$}}
\\
\\
\\
\subsection{$\mu$ = 4133 bytes, $\sigma$ = 15.8 bytes}

\begin{figure}[H]
\centerline{\includegraphics[scale = 1.0]{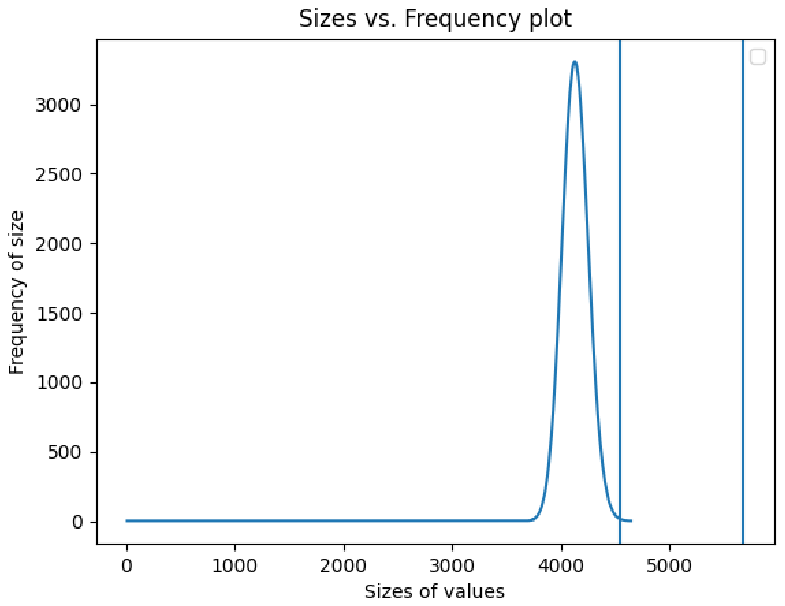}}
\caption{Old configuration of Slab Class chunk sizes. Each vertical line represents a slab class' chunk size}
\label{fig4.1}
\end{figure}

\begin{table}[H]
\caption{$\mu$ = 4133 bytes, $\sigma$ = 15.8 bytes}
\begin{center}
\scalebox{0.9}{
\begin{tabular}{|c|c|c|c|}
\hline
\textbf{Measurement Metric}&{\textbf{Old Configuration}}&{\textbf{New Configuration}}\\
\hline
Available Chunk Sizes & [4544,5680] & [4246,4644]\\
\hline
Memory wasted (bytes) & 410,568,873 & 181,599,689\\
\hline
\end{tabular}}
\label{tab4}
\end{center}
\end{table}

\begin{figure}[H]
\centerline{\includegraphics[scale = 1.0]{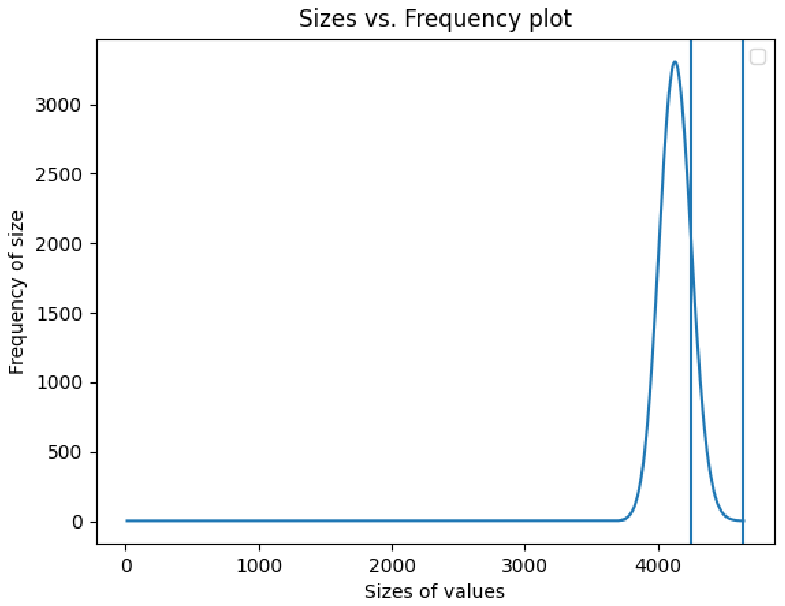}}
\caption{New configuration of Slab Class chunk sizes. Each vertical line represents a slab class' chunk size}
\label{fig4.2}
\end{figure}

{\textbf{The percentage of wasted memory recovered in the above case is 55.76$\%$}}
\\
\\
\\
\subsection{$\mu$ = 8131 bytes, $\sigma$ = 15.2 bytes}
\begin{figure}[H]
\centerline{\includegraphics[scale = 1.0]{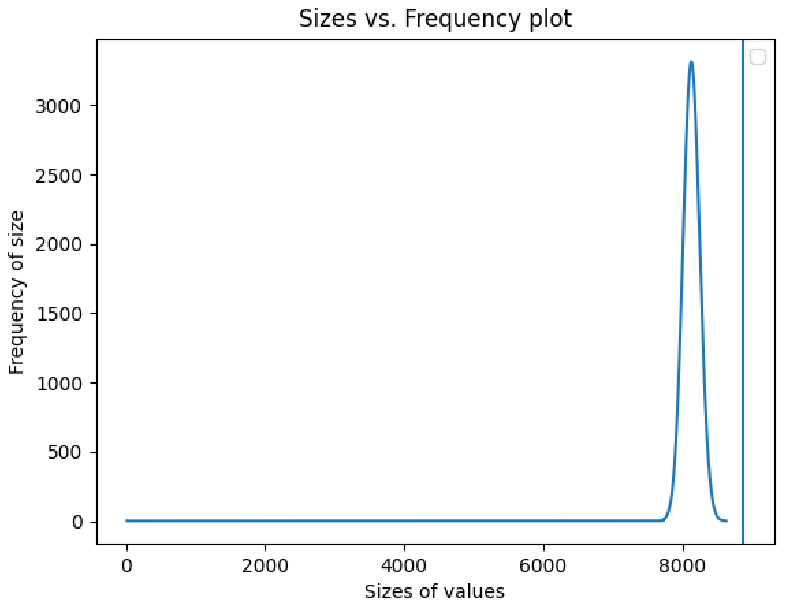}}
\caption{Old configuration of Slab Class chunk sizes. Each vertical line represents a slab class' chunk size}
\label{fig5.1}
\end{figure}

\begin{table}[H]
\caption{$\mu$ = 8131 bytes, $\sigma$ = 15.2 bytes}
\begin{center}
\scalebox{0.9}{
\begin{tabular}{|c|c|c|c|}
\hline
\textbf{Measurement Metric}&{\textbf{Old Configuration}}&{\textbf{New Configuration}}\\
\hline
Available Chunk Sizes & [8880] & [8628]\\
\hline
Memory wasted (bytes) & 748,193,597 & 496,353,869\\
\hline
\end{tabular}}
\label{tab5}
\end{center}
\end{table}

\begin{figure}[H]
\centerline{\includegraphics[scale = 1.0]{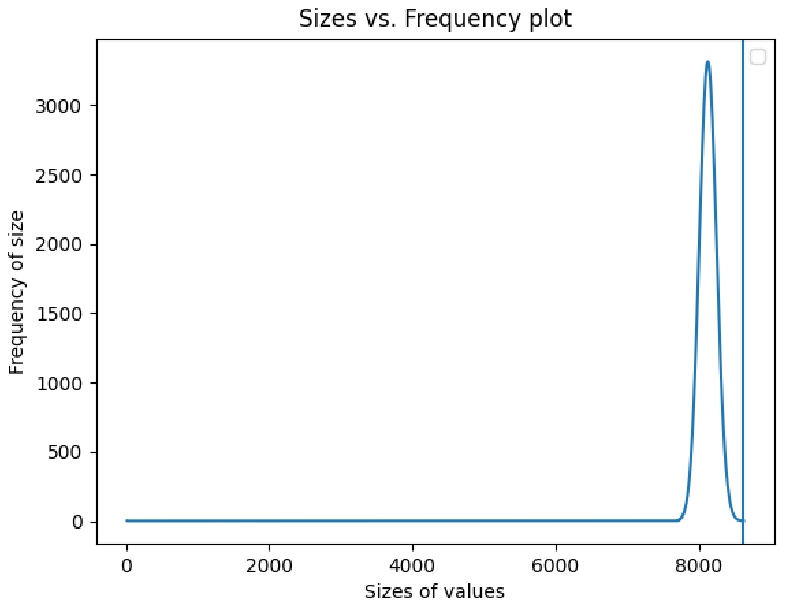}}
\caption{New configuration of Slab Class chunk sizes. Each vertical line represents a slab class' chunk size}
\label{fig5.2}
\end{figure}

{\textbf{The percentage of wasted memory recovered in the above case is 33.65$\%$}}
\\
\\
\\
\section{Discussion}
\subsection{Best and Worst case scenarios in terms of savings achieved by our algorithm}
In the best-case scenario, all of the entries into the Memcached server are of the same size, or, the number of different sizes utilized is lesser than the number of slab classes initialized. This would result in our algorithm achieving 100$\%$ storage efficiency. 

In the worst-case scenario, the data is distributed in such a manner that the default configuration happens to be the most efficient one. Here, running our algorithm would not affect storage efficiency. Two possibilities that could lead to this:
\begin{enumerate}
    \item All entered items' sizes coincide exactly with the chunk sizes generated by Memcached by default using its geometrically progressing sequence of chunk sizes. 
    \item The probability distribution of the frequency of the item for a given size follows an exponentially decreasing pattern proportional to 1.25\textsuperscript{-n}; here, the default configuration will once again fit the data in the most optimal configuration.
\end{enumerate}
However, it should be noted that it is highly improbable that either scenario arise in the real-world.

\subsection{Real world impact}
Using the above solution in a large scale environment could help servers recover terabytes of memory, increase their efficiency, and subsequently save on both power and time due to decreased RAM needs.
In 2008, Facebook's Memcached servers had 28TB of RAM allocated to instances of Memcached alone \cite{b6}. Assuming that Facebook servers also follow a similar trend of roughly 10$\%$ total Memcached memory wastage, if we replicate the results of normal distributions on their servers, i.e cutting their wastage by approximately 50$\%$, we could see a memory savings of over 1 Terabyte on Facebook servers alone. In general, we could achieve savings in memory costs of approximately 5\% for an ordinary user.

\subsection{Convergence of Algorithm}
Our greedy algorithm appears to converge to a Global minimum, i.e, whatever result it finally produces will have to be the best possible result, unlike other greedy solutions which tend to converge to local minimums. There are two reasons to conclude this. Firstly, directly neighboring configurations can result in improved performance, only if there is scope for one. Moving a slab class from a region of better efficiency to worse efficiency can never result in a better result in the long run. Secondly, the algorithm was shown to converge to the same minimum even after running a hundred restarts on the most varied distribution in all of the tests.

\subsection{Influence of standard deviation}
Data with a lower standard deviation can cause more significant savings on memory. Due to slab classes being able to shift to regions with a higher frequency of values, a lower standard deviation would mean that slab classes can crowd around a smaller region and therefore reduce the wastage in memory.

\section{Conclusion}

Memory holes are a problem that has long plagued Memcached. By using a greedy algorithm to alleviate memory holes in Memcached while using the predefined number of slab classes, we preserve Memcached's characteristic speed in storing and retrieving values stored inside it along with increasing its memory efficiency and not drastically affecting eviction rates.

Further work on this topic could investigate the effect of increasing the number of slab classes on the latency of Memcached and weigh the benefits of the increase in memory storage efficacy against the deterioration of the latency and eviction rates of the Memcached server. A more thorough analysis of this trade-off could potentially help make Memcached an even more efficient solution.

\section*{Acknowledgments}

The authors of the paper thank Dormando for providing insights on understanding how to modify slab classes in Memcached.

The authors would also like to thank Nithin Thekkupadam Narayanan, Graduate Student at MIT; and Aravind Asokan, Sr. UX/HMI Engineer, MIT SM '16 for their comments on earlier versions of this manuscript.

\end{document}